\documentclass{jaa}
\usepackage{natbib}

\usepackage{graphicx}
\usepackage{dcolumn}
\usepackage{bm}

\usepackage{aasmacros}

\usepackage[table,xcdraw,svgnames]{xcolor}
\usepackage[colorlinks]{hyperref}
\hypersetup{
    citecolor=Blue,
    linkcolor=Red,   
    urlcolor=Blue}
    
\begin{document}\sloppy

\title{Wave Effects in Double-Plane Lensing}

\author{RAHUL RAMESH\textsuperscript{1}, 
ASHISH KUMAR MEENA\textsuperscript{1,2} and 
JASJEET SINGH BAGLA\textsuperscript{1*}}

\affilOne{$^1$Department of Physical Sciences, IISER Mohali, Sector 81,
  SAS Nagar, Punjab, India - 140306.\\}
\affilTwo{$^2$Physics Department, Ben-Gurion University of the Negev, 
   P.O. Box 653, Be'er-Sheva 8410501, Israel}

\twocolumn[{

\maketitle
\vspace{1em}
\corres{jasjeet@iisermohali.ac.in}

\msinfo{xx yy 2021}{xx yy 2021}

\begin{abstract}
  We discuss the wave optical effects in gravitational lens systems
  with two point mass lenses in two different lens planes. We identify
  and vary parameters (i.e., lens masses, related distances, and their
  alignments) related to the lens system to investigate their effects
  on the amplification factor. We find that due to a large number of
  parameters, it is not possible to make generalized statements
  regarding the amplification factor.  We conclude by noting that the
  best approach to study two-plane and multi-plane lensing is to study
  various possible lens systems case by case in order to explore the
  possibilities in the parameter space instead of hoping to generalize
  the results of a few test cases. We present a preliminary analysis
  of the parameter space for a two-lens system here. 
\end{abstract}


}]

\doinum{12.3456/s78910-011-012-3}
\artcitid{\#\#\#\#}
\volnum{000}
\year{0000}
\pgrange{1--}
\setcounter{page}{1}
\lp{1}

\section{Introduction}
\label{sec:introduction}
In the last few decades, gravitational lensing has become a useful 
tool to probe properties of the Universe~\citep{2010CQGra..27w3001B,
  2011A&ARv..19...47K, 2020A&ARv..28....7U}.  
However, most of its applications are discussed in the context of
gravitational lensing of Electromagnetic (EM) radiation originating
from a distant source.  
Recent detection of gravitational wave (GW) signals from merging
compact objects by Laser Interferometer Gravitational-Wave
Observatory~\citep[LIGO;][]{2019PhRvX...9c1040A, 2020arXiv201014527A}
has opened a new avenue for application of gravitational
lensing. 

GWs are subject to deflection due to intervening matter as they travel
from the source to the observer \citep{1974IJTP....9..425O,
  1999PhRvD..59h3001B} in a manner similar to EM waves. 
As the GW signal in these direct observations originates from sources
at cosmological distances, the possibility of strong lensing due to an
intervening galaxy or galaxy cluster is significant
\citep{2018MNRAS.476.2220L, 2018IAUS..338...98S, 2019arXiv190103190B,
  2018arXiv180205273B, 2020arXiv200613219B}. 
For gravitational lensing of GW signals in the LIGO frequency range,
lensing due to a galaxy or a galaxy cluster can be described in the
same manner as that of EM radiation, i.e., in the geometric optics
limit.  
This is possible as the Schwarzschild radius of the lens is much
larger in comparison to the wavelength of the GW signal ($R_{\rm Sch}
\gg \lambda_{\rm GW}$). 
However, the same is not true for gravitational lensing of GWs (in
LIGO frequency range) due to stellar mass objects, as the wavelength
of GW signal in this case is comparable to the Schwarzschild radius of
the lens ($R_{\rm Sch} \sim \lambda_{\rm GW}$). 
As a result, wave effects becomes significant, and this leads to frequency
dependent amplification of GWs.
These wave effects due to an isolated single/double point mass lens 
\citep[e.g.,][]{2003ApJ...595.1039T, 2018PhRvD..98j3022C} 
or a point mass with external effects
\citep[e.g.,][]{2019A&A...627A.130D, 2020MNRAS.492.1127M,
  2021arXiv210203946M} have been discussed in recent studies and it
has been shown that if the point mass lens is embedded in an external
environment then the wave effects may obtain a significant boost.

In the above studies, only a single lens plane was considered.
Another interesting scenario is the lensing due to a double plane
lens. \citet{1984ApJ...276..440S} were amongst the first to explore the framework of double
plane lensing due to galaxy scale lenses, and \citet{1987MNRAS.224..283S} applied the formalism to consider dual dark matter halos as double plane lenses.
\citet{1988MNRAS.235.1073K} later pointed out that the population of double plane lenses is small, and is only expected to be approximately $1\%-10\%$ of
all lensed systems. 
Although the gravitational lensing of gravitational waves due to
galaxy scale lenses in double plane configuration can be studied using
the conventional geometric approach, the microlenses residing in these
lens galaxies can lead to wave effects. 
As these microlenses are lying in two different lens planes, the
corresponding wave effects can be significantly different from the
case of one or two-point mass lenses in single plane.
\citet{1993A&A...268..453E} presented a detailed study of two-point
mass lens in double plane lensing in the geometric optics limit and
pointed out various differences between single and double-plane
lensing. 

In the present work, we focus on the wave effects in double plane
lensing due to two point mass lenses. 
While we consider both the point lenses as isolated lenses for the bulk of the paper, we also briefly explore cases wherein these lenses are embedded in foreign galaxies.
The method to calculate the amplification factor in a generalized
N-plane lensing scenario is presented in \citet{2003astro.ph..9696Y}. 
For reference, we also briefly explore the one and two-point 
mass single plane lenses.
To the best of our knowledge, the dependence of the amplification
factor of the two-point mass single (or double) plane lens on its various
parameters has not been studied in as much detail as what we present.  

The rest of the paper is organised as follows: In
\S\ref{sec:basic_lensing}, we review the basics of geometric and wave
optics in gravitational lensing.  
In \S\ref{sec:1plane}, we revisit the single and double point mass lens in
a single plane.
In \S\ref{sec:2plane}, we explore the double plane lensing due to two point 
masses in detail: this is the core part of our work. 
Summary and conclusions are presented in \S\ref{sec:conclusions}.

\section{Gravitational Lensing}
\label{sec:basic_lensing}

In this section, we briefly review the relevant basics of
gravitational lensing in the geometric optics
regime~\citep{1992grle.book.....S} and wave optics
regime~\citep{2003ApJ...595.1039T} for a single lens plane. 
The gravitational lensing of a distant source by an intermediate lens
mass distribution (between source and observer) can be described by the
so-called gravitational lens equation,
\begin{equation}
    \bm{y} = \bm{x} - \bm{\alpha}(\bm{x}),
    \label{eq:lens_equation}
\end{equation}
where $\mathbf{x}=\bm{\xi}/\xi_0$, $\mathbf{y}=\bm{\eta}D_d/\xi_0D_s$
are the dimensionless source and image positions in the source and
lens plane, respectively, and $\xi_0$ is an arbitrary length scale.  
$D_{\rm d}$, $D_{\rm s}$ and $D_{\rm ds}$ are the angular diameter
distances from observer to lens, observer to source and from lens to
source, respectively. 
$\bm{\alpha}(\bm{x})$ is the scaled deflection angle and related to
the lens potential as: $\bm{\alpha}(\bm{x}) = \nabla \psi(\bm{x})$. 
The projected lens potential is given as 
\begin{equation}
    \psi(\bm{x}) = \frac{1}{\pi} \int d^2 x' \kappa(\bm{x}') \ln
    |\bm{x} - \bm{x}'| 
    \label{eq:lens_potential},
\end{equation}
with
\begin{equation}
    \kappa(\bm{x}) = \frac{\Sigma(\bm{x})}{\Sigma_{\rm cr}}, \quad
    \Sigma_{cr} = \frac{c^2}{4\pi G}\frac{D_{\rm s}}{D_{\rm d} D_{\rm ds}},
    \label{eq:convergence}
\end{equation}
where $\kappa$ is known as \textit{convergence} and represents the
dimensionless surface mass density of the lens.

For a lensed signal coming from the source, the corresponding time
delay ($t_{\rm d}$) with respect to the unlensed signal is given by
\begin{equation}
    t_{\rm d} = \frac{\xi^2_0}{c}\frac{D_{\rm s}}{D_{\rm d} D_{\rm ds}}(1+z_{\rm d})
    \left[\frac{(\bm{x}-\bm{y})^2}{2}-\psi(\bm{x})+\phi_m(\bm{y})\right],
    \label{eq:time_delay}
\end{equation}
where $z_d$ is the lens redshift, $\xi_0$ is an arbitrary length scale
used to make Equation~\ref{eq:lens_equation} dimensionless and
$\phi_m(\bm{y})$ is a constant which is independent of lens properties
and can be chosen to simplify the calculations.  

The above mentioned formalism of gravitational lensing is valid as
long as the Schwarzschild radius of the lens is much greater than the
wavelength of the signal, i.e., $R_{\rm Sch} \gg \lambda$. 
Hence, it is applicable in case of gravitational lensing of EM waves 
originating from a distance source due to a lens in the mass range
from stellar mass objects ($\sim 1{\rm M}_\odot$) to cluster of
galaxies ($\sim10^{15}{\rm M}_\odot$). 
On the other hand, if the wavelength of the signal is of the order of
the Schwarzschild radius of the lens, then the framework of geometric
optics no longer holds and we need to take wave effects into account. 
For light with wavelength $\sim 1\mu {\rm m }$, wave effects are
considerable if lens mass is $\sim 10^{-9} {\rm M}_\odot$.

For gravitational waves in the LIGO frequency range (10 Hz-$10^4$ Hz),
gravitational lensing due to a galaxy or a cluster of galaxies can be
well described using the geometric optics approach outlined above.
However, wave effects arise if the signal is lensed by stellar mass 
objects  ($10{\rm M}_\odot - 10^4 {\rm M}_\odot$; see figure 1 in
\citealt{2020MNRAS.492.1127M}) as the corresponding Schwarzschild
radius ($R_{\rm Sch}$) is of the order of the wavelength of the signal
($\lambda_{\rm GW}$). 

Thus, in a typical scenario of microlensing of gravitational waves,
one cannot use geometric optics as the wave effects are not
negligible: these manifest as a frequency dependent amplification
factor and phase shift. 
The corresponding amplification factor which is defined as the ratio
of the lensed and unlensed signals is given
as~\citep{1999PThPS.133..137N, 2003ApJ...595.1039T} 
\begin{equation}
    F\left(f,\mathbf{y}\right)=\frac{D_{\rm s} \: \xi_0^2
      \left(1+z_{\rm d}\right)} 
    {c D_{\rm d} D_{\rm ds}}\frac{f}{\textit{i}} \int d^2\mathbf{x} \:
    \exp\left[2\pi\textit{i} f t_d\left(\mathbf{x},\mathbf{y}\right)\right], 
\label{eq:amplification_factor}
\end{equation}
where $f$ is the frequency of the gravitational wave signal, and the
rest of the symbols have their usual meaning.
One can see that the amplification factor depends on the frequency of
the GW signal. 
Hence, different frequency components of the signal are modulated by
varying factors, unlike the case of achromatic lensing in the regime
of geometric optics. 
As the amplification factor, $F(f,\bm{y})$, is a complex quantity,
wave effects modify both the amplitude and the phase of the GW
signal. 

As we transition towards high frequencies (geometric optics regime),
the integral in the above equation turns highly oscillatory and only
the stationary points of the time delay ($t_{\rm d}$) contribute
significantly. 
The form of Equation~\ref{eq:amplification_factor} in geometric optics
is 
\begin{equation}
    F\left(f,\mathbf{y}\right)=\sum_i\sqrt{|\mu_j|}\exp\left(2\pi
    \textit{i} f t_{d,j}-\textit{i}\pi n_j\right),
\end{equation}
where $t_{d,j}$ is the value of time delay for $j$-th image, $\mu_j$
is the amplification of the $j$-th image and $n_j$ is the Morse index
with values 0, 1/2, 1 for images corresponding to minima, saddle and
maxima of the time delay function, respectively.
One can see that phase of the lensed GW signal corresponding to saddle
(maxima) gets modified by $e^{-i\pi/2}(e^{-i\pi})$. 

\section{Single Plane Lensing}
\label{sec:1plane}

\subsection{Single Plane One-Point Mass lens}
\label{ssec:1plane_1mass}

Although one cannot solve Equation~\ref{eq:amplification_factor}
analytically for general lens mass models, the isolated point mass
lens model has an analytical solution \citep{1974PhRvD...9.2207P}:
\begin{eqnarray}
    F\left(\omega,y\right) &=& \exp\left[\frac{\pi
                           \omega}{4}+\frac{\textit{i}\omega}{2}
                           \Bigg\{\ln\left(\frac{\omega}{2}\right) -
                           2\phi_m\left(y\right)\Bigg\} 
                           \right]\nonumber\\  
&&
   \Gamma\left(1-\frac{\textit{i}\omega}{2}\right){}_{1}F_{1}
   \left(\frac{\textit{i}\omega}{2},1;\frac{\textit{i}\omega }{2}y^2
   \right),  
\label{eq:point_wave}
\end{eqnarray}
where $\omega=8\pi G (1+z_{\rm d})M_{\rm L}f/c^3$, $y = |\mathbf{y}|$,
$\phi_m\left(y\right) 
= \left(x_{\rm m}-y\right)^2/2-\ln\:x_{\rm m}$ with 
$x_{\rm m}=\left(y+\sqrt{y^2+4}\right)/2$.
Here we have used the Einstein radius of the lens system as the
relevant scale length: $\xi_0=\left(4GM_{\rm L} D_{\rm d} D_{\rm
  ds}/c^2 D_{\rm s}\right)^{1/2}$. 

\begin{figure*}[ht!]
  \centering
  \includegraphics[width=16.6cm, height=12cm]{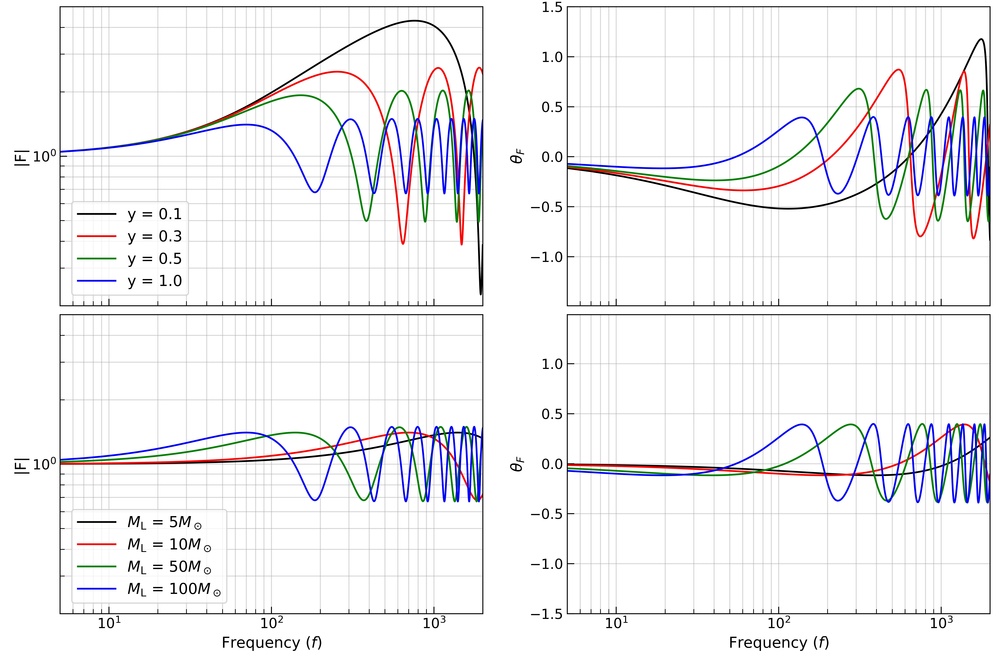}
  \caption{Gravitational lensing due to point mass lens: The top row
    represents the amplification factor ($|F|$) and phase shift
    ($\theta_F$) due to a point mass lens of 100$M_\odot$ with
    different source positions ($y$) in left and right panel,
    respectively. Similarly, the bottom row shows the amplification
    factor and phase shift factor due to a point lens with different
    lens mass values  and source position, $y=1$.}
  \label{fig:point_lens}
\end{figure*}

The amplification factor of an isolated point mass lens depends on two
parameters: the lens mass $M_{\rm L}$, and the source position ($y$)
in the source plane. 
Figure~\ref{fig:point_lens} shows the dependency of $F(f)$ on these
two parameters. 
The top row represents the absolute value of the amplification factor
($|F|$) and the phase shift ($\theta_F$) in left and right panel,
respectively, for different source positions ($y$). 
Here the lens mass ($M_{\rm L}$) has been fixed at $100M_\odot$.
Similarly, the bottom row represents the variation in the lens mass
$M_{\rm L}$ with a fixed source position $y=1$.
One can see that for a fixed lens mass, variations in the source
position changes the amplitude of the oscillations. 
As we move the source towards the center, the amplitude of the
oscillations increases.
However, the frequency of the oscillations decreases.
On the other hand, the amplitude of the oscillations is fixed if we
fix the source position, and only the rate of oscillations change as
we vary the lens mass. 

\subsection{Single Plane Two-Point Mass lens}
\label{ssec:1plane_2mass}

In a typical galaxy, approximately half of the stellar mass objects
are in binary systems.
Hence, a two-point (binary) mass lens is a natural generalization of
the isolated  point mass lens~\citep{1986A&A...164..237S}.
The lensing potential corresponding to a two-point mass lens is given as
\begin{equation}
  \psi\left(x_1,x_2\right) = \mu_1 \ln\left(\bm{x}-\bm{{\rm L}}\right) 
  + \mu_2 \ln\left(\bm{x}+\bm{{\rm L}}\right),
  \label{eq:binary_potential}
\end{equation}
where $\left(\mu_1, \mu_2\right) = 
\left(M_1/M_{\rm T}, M_2/M_{\rm T}\right)$, with $M_{\rm T} = M_1 + M_2$.
For a two-point mass lens, one can always choose a coordinate system
in which both point masses lie on the x-axis and are at equal distance
from the center.  
Hence, the distance vectors for the first and second point masses are
$({\rm L}, 0)$ and $(-{\rm L}, 0)$, respectively. 

\begin{figure*}[ht!]
  \centering
  \includegraphics[width=16.6cm, height=6cm]{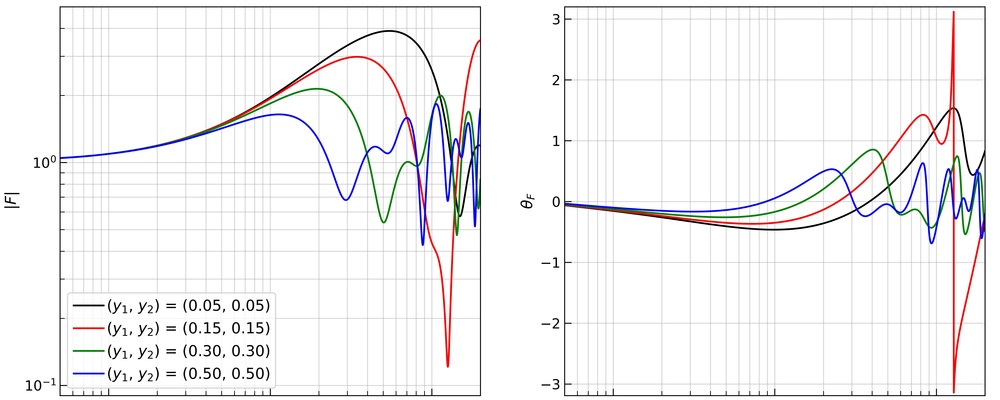}
  \includegraphics[width=16.6cm, height=6cm]{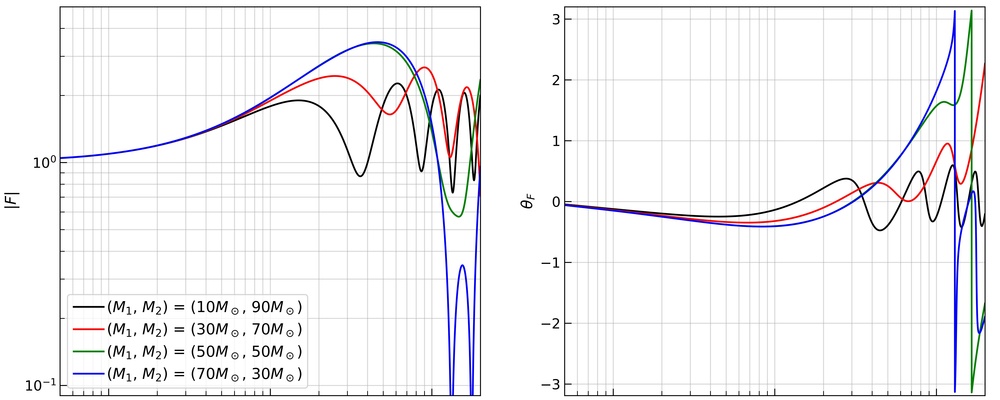}
  \includegraphics[width=16.6cm, height=6.5cm]{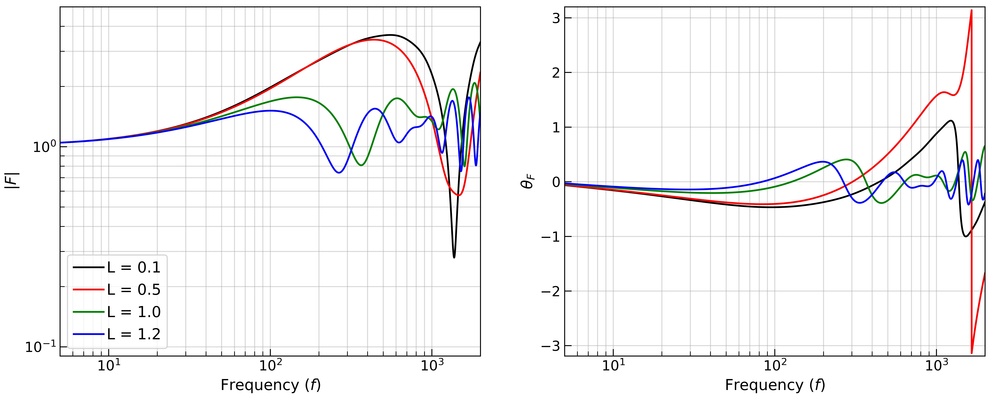}
  \caption{Gravitational lensing due to binary lens: The top row
    represents the amplification factor ($|F|$) and phase shift
    ($\theta_F$) due to a binary lens with binary lens masses of
    ($M_1$, $M_2$) = (50$M_\odot$, 50$M_\odot$) at a separation
    ($\bm{{\rm L}}$) = 0.5. The different values of ($y_1$, $y_2$) are
    mentioned in the left panel.  The middle row represents the
    variation in lens masses while fixing the source position to
    ($y_1$, $y_2$) = (0.1, 0.1) and at a separation of ($\bm{{\rm
        L}}$) = 0.5.  The bottom row represents the variation in the
    separation vector while fixing the source position to ($y_1$,
    $y_2$) = (0.1, 0.1) and binary lens masses ($M_1$, $M_2$) =
    (50$M_\odot$, 50$M_\odot$).} 
  \label{fig:binary_lens_wave}
\end{figure*}

A two-point mass lens system can be described by a total of five
parameters, namely, the source position ($y_1,y_2$), the lens masses
($M_1,M_2$), and their separation 2L. 
Depending on the values of these parameters, one obtains either a
three or a five-image configuration.
Gravitational lensing due to a binary lens in geometric optics has
already been studied in great detail
\citep[e.g.,][]{1986A&A...164..237S, 1993JMP....34.4093W,
  2003PThPh.110..425A, 2009ApJ...690.1772P, 2020Univ....6..106B}. 
However, the same is not true for the wave optics regime.

Figure~\ref{fig:binary_lens_wave} shows the $F(f)$ curves for a binary
lens with different values of the above mentioned parameters. 
The corresponding caustics and critical line plots are shown in 
Figure~\ref{fig:binary_lens_geom} in \ref{sec:appendix_A}.
The top panel in Figure~\ref{fig:binary_lens_wave} shows the
amplitude ($|F|$) and phase shift ($\theta_F$) in the left and right
panel, respectively, for different source positions keeping the lens
masses and the separation vector fixed.  
Here, we fix lens masses ($M_1,M_2$) and the separation vector
($\bm{{\rm L}}$) to be ($50M_\odot, 50M_\odot$) and (0.5, 0),
respectively. 
Analogous to a single point mass lens, as we move towards the center,
the amplitude of the fluctuations in amplitude and phase shift
increases and the frequency of the oscillations decreases. 
However, the presence of extra images due to the binary lens 
introduces further fluctuations in the amplitude and phase shift.
Instead of the source position, if we vary the mass of the two
components of the binary (centre panel), at low frequency ($f<100$Hz),
we make observations similar to the previous case (and the isolated
point mass lens).  
However, at $f>100$Hz, we notice differences due to the varying number
of images, the time delay between these images, and the magnification
of the various images.  
This can be seen in the green and blue curves which correspond to
($M_1,M_2$) = ($50M_\odot, 50M_\odot$) and ($70M_\odot, 30M_\odot$). 
Both these configurations give rise to a five-image geometry (centre
panel of Figure~\ref{fig:binary_lens_geom}).  
However, the blue curve shows significant de-amplification compared to
the one in green. 
In the bottom panel, we vary the separation between the two lenses by
fixing the source position to (0.1, 0.1) and lens masses to 
($50M_\odot, 50M_\odot$). 
For the case of small values of ${\rm L}$, we converge towards
the point mass lens as expected (${\rm L}$ = 0.1; black curve).
Increasing the separation leads to different types of interference
patterns depending on the number of images and the corresponding time
delays. 
 
\section{Double-Plane Lensing}
\label{sec:2plane}

The presence of a second lens plane contributes an additional
deflection angle in the gravitational lens equation (in thin lens
approximation) given as (in dimensionless form;
\citealt{1992grle.book.....S}), 
\begin{equation}
  \bm{y} = \bm{x}_1 - \bm{\alpha}_1(\bm{x}_1) - \bm{\alpha}_2(\bm{x}_2),
  \label{eq:lens_equation_2plane}
\end{equation}
where $\bm{y}$ is the unlensed source position in the source plane. 
$\bm{\alpha}_1 (\bm{\alpha}_2)$  is the scaled deflection angle and 
$\bm{x}_1 (\bm{x}_2)$ is the corresponding impact parameter in the first 
(second) lens plane, respectively. 
Due to the additional deflection (introduced by the second lens plane), 
the gravitational lens equation no longer remains a gradient mapping from
image to source plane, leading to new interesting image properties (in 
geometric optics approximation).
The time delay function in case of double-plane lensing is given as
\begin{equation}
    t_{\rm d}(\bm{x}_1, \bm{x}_2, \bm{y}) = t_{12}(\bm{x}_1, \bm{x}_2) 
    + t_{23}(\bm{x}_2, \bm{y})
    \label{eq:time_delay_2plane},
\end{equation}
and
\begin{equation}
    t_{ij}(\bm{x}_i, \bm{x}_j) = \frac{1+z_i}{c} \frac{D_i D_j}{D_{ij}} \left[
    \frac{(\bm{x}_i - \bm{x}_j)^2}{2}-  \beta_{ij} \: \psi_1(\bm{x}_i) \right],
\end{equation}
where $\beta_{ij} = D_{ij} D_s/D_j D_{is}$ is a combination of various angular
diameter distances related to the lens system with the indices $i,j=1,2,3, \: 
i<j$ and $j=3$ represents the source plane.
The rest of the symbols have their usual meaning.

\begin{figure*}[ht!]
    \centering
    \includegraphics[width=16.6cm, height=6.0cm]{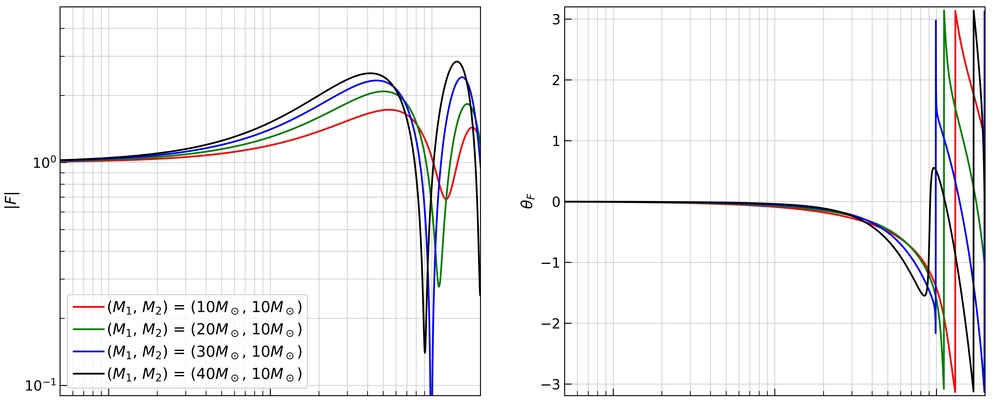}
    \includegraphics[width=16.6cm, height=6.0cm]{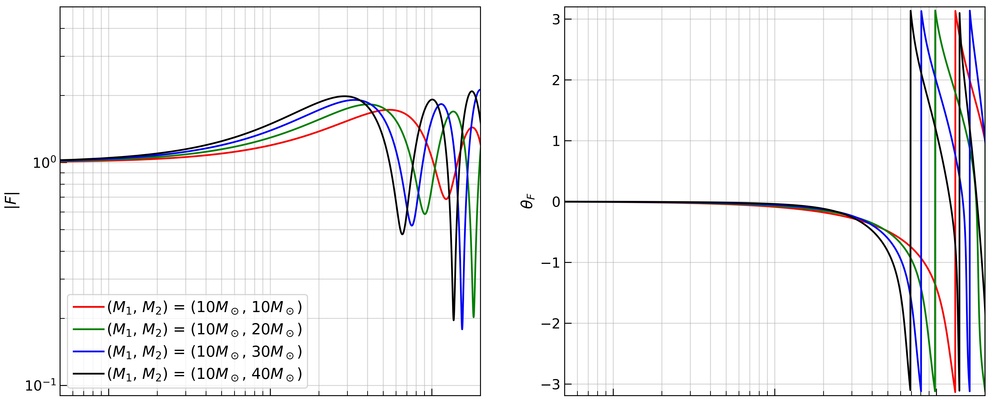}
    \includegraphics[width=16.6cm, height=6.5cm]{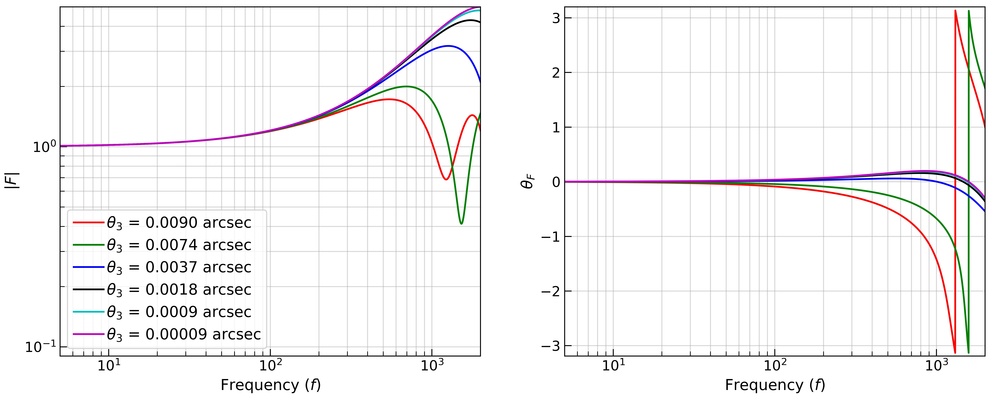}
    \caption{Dependence of the amplification factor (absolute value on the 
    left, and phase ($\Delta \Phi = - i \ln[F/|F|]$) on the right) on various 
    parameters for the special case (Equation~\ref{eq:special_case}). We vary 
    one parameter at a time, while keeping the others fixed (with values as 
    explained in detail in the main text). Top panel: As $M_1$ increases, 
    oscillations become more rapid, and the maximum value of amplification 
    increases as well. Centre panel: Similar to the previous case, oscillations 
    become more rapid, but there is no rise in value of amplification. Also, 
    oscillations here are more rapid in comparison to the previous case. 
    Bottom panel: Angular position of the observer is varied. When the observer 
    is closer to the line joining the source with the two lenses, oscillations 
    are slower, but the amplification is larger.  As $\theta_3$
    reduces, we also note that the curves tend to converge.} 
    \label{fig:multi_plane_special_case}
\end{figure*}

\begin{figure*}[ht!]
    \centering
    \includegraphics[width=16.6cm, height=6.0cm]{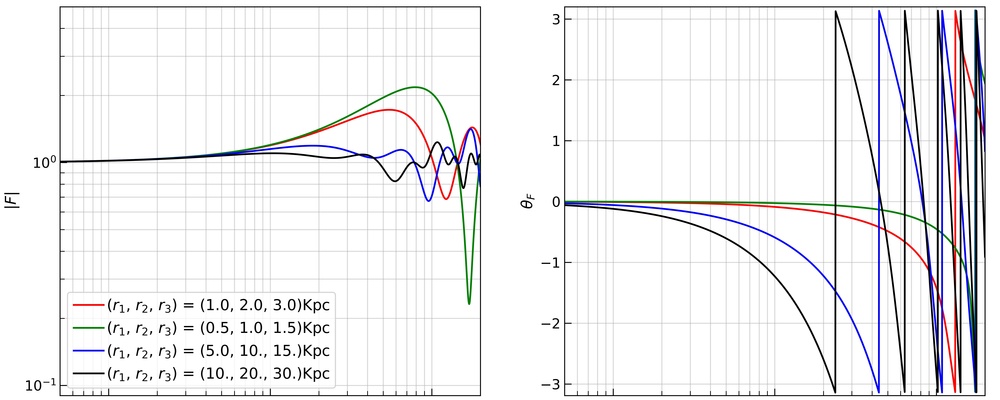}
    \includegraphics[width=16.6cm, height=6.5cm]{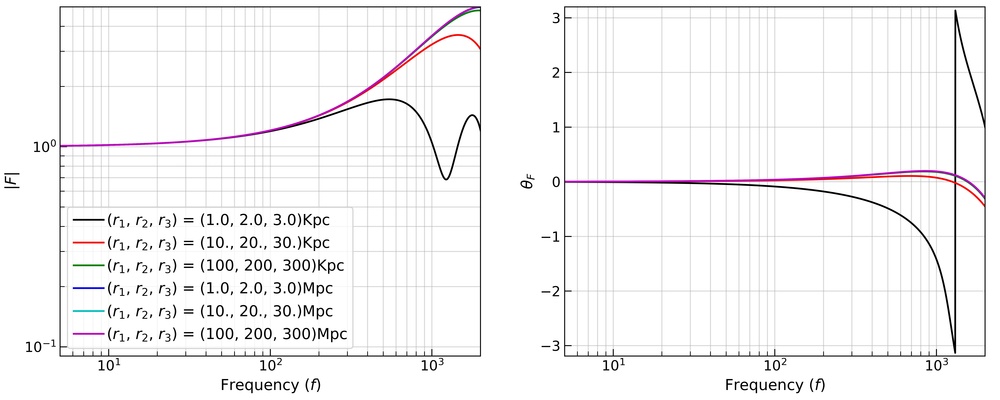}
    \caption{Dependence of the amplification factor on the radial distances of the three objects for the special case (Equation~\ref{eq:special_case}). While there are multiple ways to vary the three distances, we do so by maintaining a constant ratio between the three values. In the top panel, we fix the angular position of the observer, while we fix the transverse distance of the observer in the bottom panel. In the former case, we note that greater values of ($r_1, r_2, r_3$) lead to an increase in rate of oscillation, coupled with a drop in the maximum value of amplification. Trends are opposite in the bottom panel.} 
    \label{fig:multi_plane_special_case_dist}
\end{figure*}

Similar to \S\ref{sec:1plane}, the above mentioned equations are valid only
within the regime of geometric optics. 
To study the wave effects in double plane lensing, one needs to generalize 
Equation \ref{eq:amplification_factor}.
Such a generalization has been presented in \citet{2003astro.ph..9696Y} using
path integral formalism developed in \citet{1999PThPS.133..137N}.
In the case of N-plane lens system, the amplification factor is given by 
\citep{2003astro.ph..9696Y}:

\begin{equation}
F\left(f,\mathbf{y}\right) = C f^N \int d^2\mathbf{x_1}\:... d^2\mathbf{x_N}
\exp\left[2\pi\textit{i} f
  t_d\left(\mathbf{x_1},...,\mathbf{x_N},\mathbf{y}\right)\right] 
\label{eq:amplification_factor_mp}
\end{equation}
where different $\mathbf{x_i}$ represent two dimensional vectors in different
lens planes.
$f$ represents the frequency of the signal.
The time delay function is given by $t_d\left(\mathbf{}x_1,...,\mathbf{x}_N,
\mathbf{y}\right) = \Sigma_{i = 1}^{N} t_{i,
  i+1}\left(\mathbf{x}_i,\mathbf{x}_{i+1}\right)$   
with $\mathbf{x}_{N+1} = \mathbf{y}$.
$C$ is a factor which depends on the various angular diameter
distances under consideration and given as
\begin{equation}
C = \left(\frac{-i}{c}\right)^N \: \prod_{i=1}^N (1+z_i) \frac{D_{0,i}
  D_{0,i+1} }{ D_{i,i+1} }, 
\end{equation}
where $z_{i}$ is the redshift for the $i$-th lens plane and $D_{0,i},
\: D_{0,i+1}$, and $D_{i,i+1}$ are the angular diameter distances
between observer and $i$-th lens plane, observer and $(i{+}1)$-th lens
plane, and $i$-th and $(i{+}1)$-th lens plane. 

In general, Equation~\ref{eq:amplification_factor_mp} needs to be
solved numerically.  
However, for a special case of double-plane lensing (N = 2), one can
obtain an analytic expression \citep{2003astro.ph..9696Y}. 
This special case is discussed in \S\ref{ssec:2plane_special}.
More general cases are discussed in \S\ref{ssec:2plane_general} and \S\ref{ssec:2plane_galaxy}.
Performing this `2N' integral (Equation
\ref{eq:amplification_factor_mp}) using standard quadrature methods is
not a very efficient process. 
Hence, for these general cases, we use a Monte Carlo integrator
provided by the \textsc{vegas} package \citep{2020arXiv200905112L}.

\subsection{The Special Case}
\label{ssec:2plane_special}

Consider two point mass lenses with masses $M_1$ and $M_2$, at
distances $r_1$ and $r_2$ (from the source) respectively, that are
co-linear with the source.  
For an observer at a distance $r_3$ and angular position $\theta_3$, the 
amplification factor is given by \citep{2003astro.ph..9696Y}:

\begin{eqnarray}
 F(k, \theta_3) &=& \exp(i\alpha) \exp\left(\frac{\pi k G M_{\rm T}}{c^2}\right)
 \Gamma\left(1-\chi_1\right) 
    \nonumber
    \\
    && 
    z \sum_{L=0}^\infty{\frac{(-i)^L}{(L!)^2}} 
    \Gamma\left(1+L-\chi_2\right) (xz)^L
    \nonumber
    \\
    &&
    {}_2F_1\left(1-\chi_1,1+L -\chi_2 ,1~; 1-z\right)
\label{eq:special_case}
\end{eqnarray}
where
\begin{align}
  M_{\rm T} & \coloneqq M_1 + M_2, \quad k \coloneqq (2 \pi f)/c,
\end{align}
\begin{align}
  \chi_1  &\coloneqq \frac{2ikGM_1}{c^2}, 
  \quad 
  \chi_2 \coloneqq \frac{2ikGM_2}{c^2},
\end{align}
\begin{align}
  z \coloneqq { \frac{r_3(r_2-r_1)}{r_2(r_3-r_1)}},
  \quad 
  x \coloneqq { \frac{kr_2r_3\theta_3^2}{2(r_3-r_2)}},
\end{align}
$\alpha$ is a real constant corresponding to a constant phase difference, 
and ${}_2F_1(a,b,c;d)$ is the Hyper-geometric function. 

For the special case under consideration, there are 
six different parameters on which the amplification factor depends, namely,
$r_1$, $r_2$, $r_3$, $\theta_3$, $M_1$ and $M_2$. 
In Figure~\ref{fig:multi_plane_special_case}, we vary these parameters one
at a time. 
For each case, the absolute value of the amplification factor is shown on 
the left, while the phase is shown on the right. 
In the top panel, we fix ($r_1, r_2, r_3$) = (1.0, 2.0, 3.0) kpc, $M_2$ = 10 
$M_\odot$, $\theta_3$ = 0.009 arc-sec, and vary $M_1$. 
As $M_1$ increases, oscillations become more rapid, and this is accompanied 
by an increase in the maximum value of the amplification factor. 
In the centre panel, we vary the value of $M_2$, while keeping $M_1$ = 10 
$M_\odot$ and other parameters fixed at the values mentioned above. 
As $M_2$ increases, we again notice that the oscillations become rapid, and 
the maximum value of the amplification factor also does increase slightly. 
However, in the case of varying $M_2$, oscillations are more rapid than in 
the case of varying $M_1$, and the maximum value of amplification does not 
change as sharply as in the case of varying $M_1$. 
For instance, the black curve in the centre left panel oscillates with a 
frequency of $\sim 500$ Hz, while the frequency of oscillation of the black 
curve in the top left panel is $\sim 700$ Hz. 
This set of observations suggest the following: the mass of the lens closer 
to the source ($M_1$) has a greater affect on the maximum value of the modulus 
of amplification, while $M_2$ is more important with respect to the rate at 
which the amplification factor oscillates. We also note the difference in 
behaviour of the phase at large $M_1$.

In the bottom panel, we vary $\theta_3$ while fixing ($r_1, r_2, r_3$)
= (1.0, 2.0, 3.0) kpc, and ($M_1, M_2$) = (10, 10) $M_\odot$. As $\theta_3$ 
decreases, i.e. as the observer approaches the line joining the source with 
the two lenses, two qualitative trends take place: oscillations become less 
rapid, and the maximum value of (modulus of) the amplification factor
increases.  
This is qualitatively similar to what one observes in the single lens
plane case.  We also note that the behaviour of the amplification factor tends 
to converge as $\theta_3$ reduces (e.g. the cyan and pink curves almost 
superimpose on one another).

In Figure~\ref{fig:multi_plane_special_case_dist}, we vary ($r_1, r_2, r_3$) while 
fixing $M_1$ = 10 $M_\odot$, $M_2$ = 10 $M_\odot$. 
In the top panel, we fix the angular position of the source ($\theta_3$ = 0.009 arcsec). 
While the three distances can be
varied in multiple ways, we do so by fixing the ratio between the
three values. As the distances get larger, oscillations are more
rapid, and the maximum value of amplification is smaller. We note the
similarity in observations with those of the bottom panel of Figure~\ref{fig:multi_plane_special_case}: increasing
$\theta_3$ while keeping the three distances fixed is analogous to
keeping $\theta_3$ fixed while increasing the three distances -- both
lead to an increase in the length of the perpendicular dropped from
the observer to the line joining the source with the two
lenses. 

Instead of fixing the value of $\theta_3$ while varying ($r_1,
r_2, r_3$), one can alternatively fix the value of the above mentioned
perpendicular distance. We show these trends in the bottom panel of
Figure~\ref{fig:multi_plane_special_case_dist} where we fix the perpendicular distance 
to be ${\sim} 4 {\times} 10^{12}$ meters. Once again, we note similarities to
the bottom panel of Figure~\ref{fig:multi_plane_special_case}: as the magnitude of 
($r_1, r_2, r_3$) increases, the angular position of the
observer is effectively reduced, and we observe that the oscillations of the
amplification factor become less  
rapid, and the maximum value of (modulus of) the amplification factor
increases. Also, the curves tend to
converge for large values of the three distances.

\begin{figure*}[ht!]
    \centering
    \includegraphics[width=16.6cm, height=6cm]{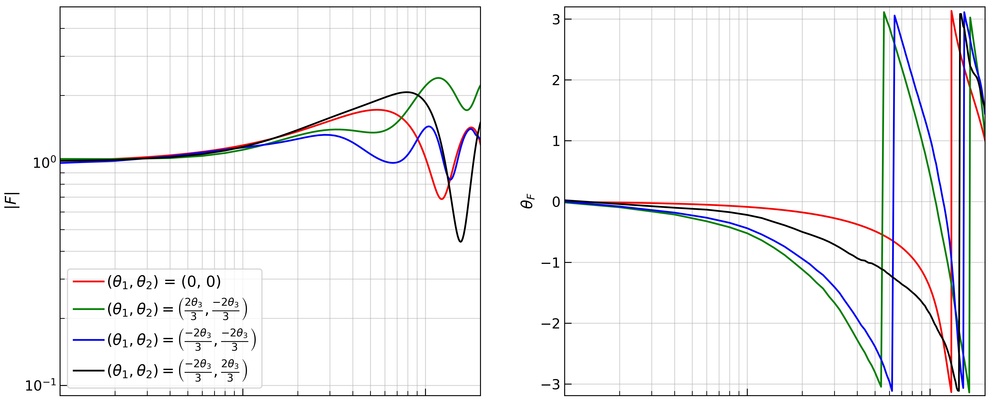}
    \includegraphics[width=16.6cm, height=6.5cm]{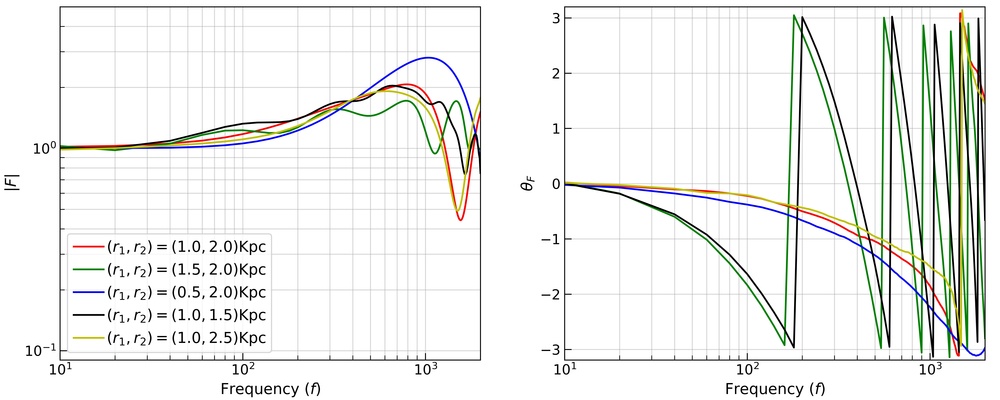}
    \caption{In-Plane Variation of angular positions -- Top Panel: Varying 
    the angular position of the two lenses. As elaborated in the main text, 
    the lens closer to the source ($M_1$) seems to have a greater effect 
    on the range of $|$F$|$ values, while the rate of oscillation of the 
    $|$F$|$ curve seems to have a greater dependence on $M_2$. Bottom 
    Panel: Varying the radial distance of lenses at fixed values of other 
    parameters. The dependence of the amplification factor on these two 
    parameters seems to be more complex than the dependence on angular
    positions.} 
    \label{fig:multi_in_plane}
\end{figure*}

\subsection{Relaxing the Co-linearity Condition}
\label{ssec:2plane_general}

In this sub-section, we relax the condition that requires the two
lenses to be co-linear with the source. 
Once we relax the same, there are two possible scenarios that one can
consider:
(1) In-Plane variation of angular positions: the source, lenses and
the observer lie along the same plane.
This introduces  
two additional parameters in the analysis, namely, $\theta_1$ and
$\theta_2$ --  the angular positions of the two lenses with respect to
the source.  
(2) Off-Plane variation of angular positions: the source, lenses and
the observer no longer lie along the same plane. In addition 
to $\theta_1$ and $\theta_2$, this case has three more parameters,
namely,  
$\phi_1, \phi_2, \phi_3$ -- the angles (measured from the source)
subtended by the  
off-plane object with respect to the plane of the paper.

\begin{figure*}[ht!]
    \centering
    \includegraphics[width=16.6cm, height=6cm]{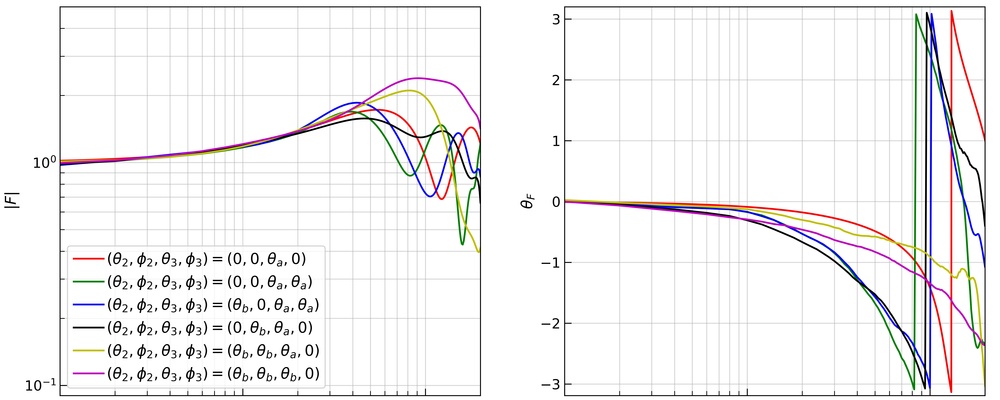}
    \includegraphics[width=16.6cm, height=6cm]{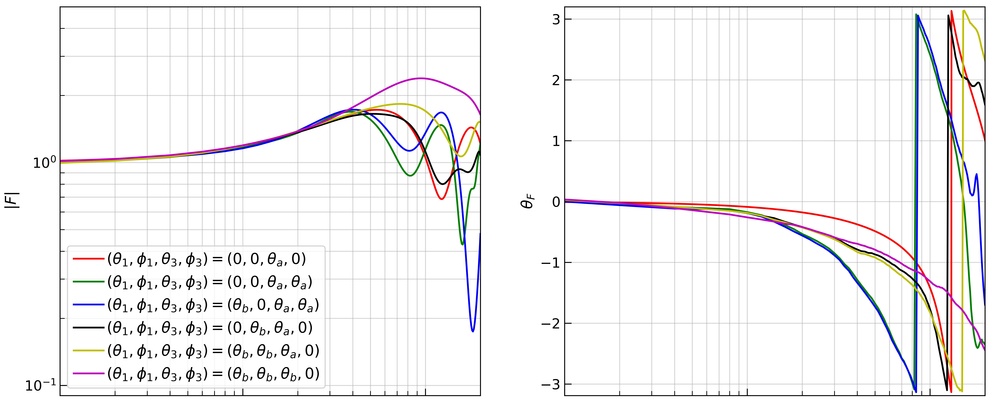}
    \includegraphics[width=16.6cm, height=6.5cm]{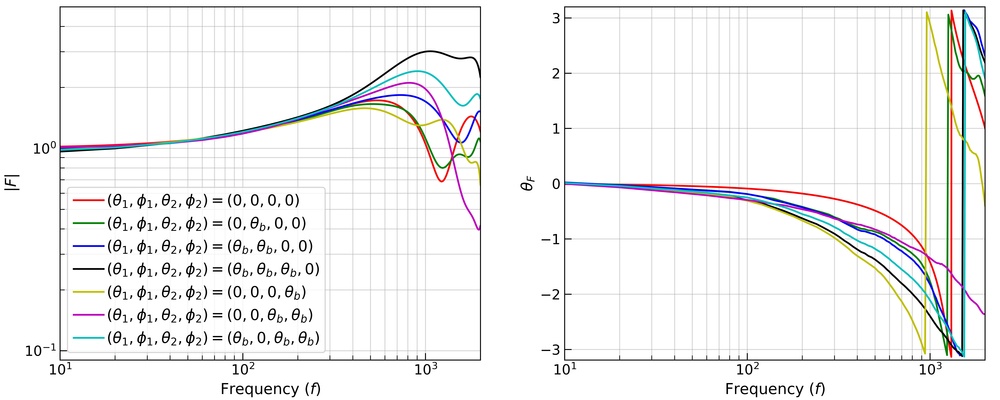}
    \caption{Off-Plane Variation of angular positions -- For simplicity, angles 
    are represented by this set of variables: ($\theta_a, \theta_b$) =
    (0.009 arc-sec, (2/3)$\theta_a$)rad.  
    As elaborated in the main text, in each panel, we choose one object (along 
    with the source) to form the primary axis of the plane under consideration. 
    One of the other two objects are displaced off-plane. We note that trends 
    are not always similar to in-plane variation of angular positions.}
    \label{fig:multi_off_plane}
\end{figure*}

We start with the In-Plane case, where the two lenses, the source and 
the observer all lie along the same plane. 
The corresponding absolute values and phase of the amplification factors are shown in
Figure \ref{fig:multi_in_plane}.
In the top panel, we choose $(r_1, r_2, r_3)$ = (1.0, 2.0, 3.0) kpc,
$(M_1, M_2)$  
= (10, 10) $M_\odot$, $\theta_3$ = 0.009 arc-sec, and vary the values of 
$(\theta_1, \theta_2)$. 
In red, for reference, we consider $(\theta_1, \theta_2)$ = (0, 0),
which is the  
same as some of the curves from Figure~\ref{fig:multi_plane_special_case}. 
For the green and blue curves, we choose $(\theta_1, \theta_2)$ =
$(\frac{2 \theta_3}{3},  
\frac{-2 \theta_3}{3})$ and $(\frac{-2 \theta_3}{3}, \frac{-2 \theta_3}{3})$. 
These two curves oscillate faster than the curve in red, with the blue curve 
portraying faster oscillations than the curve in green: oscillations are most 
rapid when both the lenses are away from the line joining the source
with the observer.  
In the curve in black, we place $M_2$ closer to the line joining the
source with  
the observer, with  $(\theta_1, \theta_2)$ = $(\frac{-2 \theta_3}{3},
\frac{2 \theta_3}{3})$.  
In this case, the curve oscillates slower in comparison to the red curve. 
While the angular positions of both lenses have an impact on both the rate 
of oscillation and the maximum value of amplification, we note the following 
difference: the angular position of $M_2$ seems to have a greater affect on 
the rate of oscillation in comparison to the maximum value of amplification. 
Comparing the blue and green curves, we note the opposite for the angular 
position of $M_1$, which seems to have a greater effect on the maximum value
reached by the amplification factor as it oscillates.

\begin{figure*}[ht!]
  \centering
  \includegraphics[width=16.6cm, height=6.0cm]{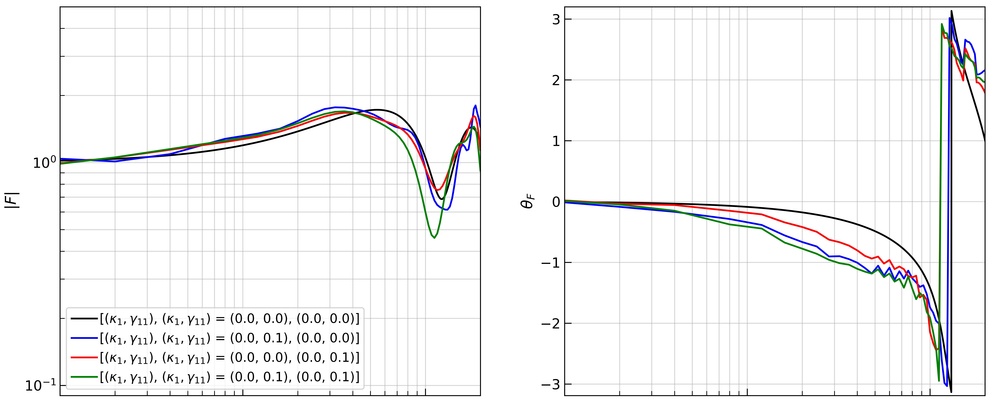}
  \includegraphics[width=16.6cm, height=6.0cm]{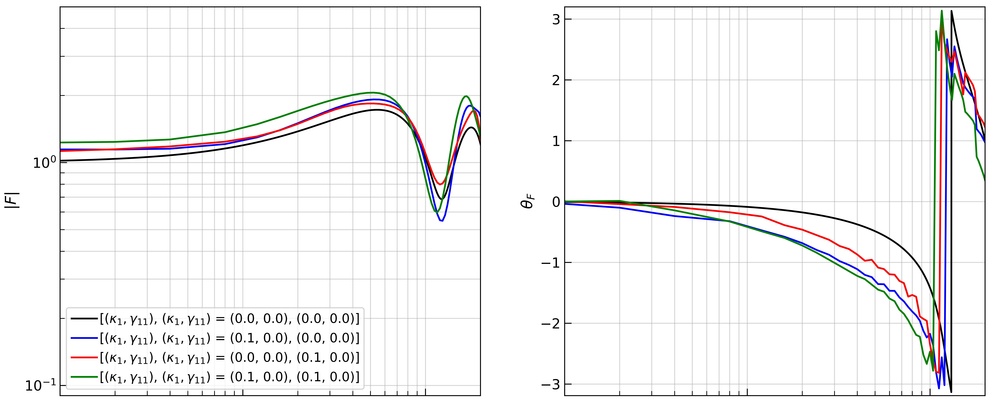}
  \includegraphics[width=16.6cm, height=6.5cm]{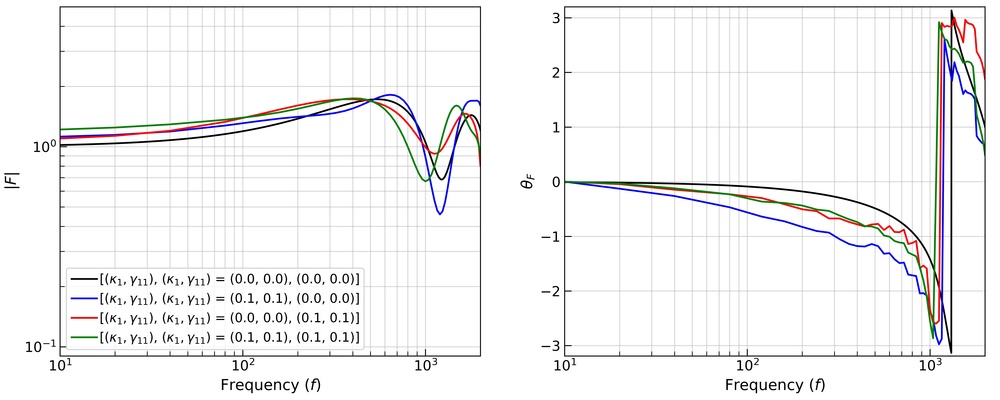}
  \caption{Variation of convergence and shear parameters for the case
    described in \S\ref{ssec:2plane_galaxy}. In the top (centre)
    panel, we examine cases with only external shear (convergence). In
    the bottom panel, we consider non-zero values of both external
    shear and convergence. External shear seems to have a greater
    effect on the rate of oscillation, while external convergence
    modifies the amplification, more so at low frequencies.} 
  \label{fig:multi_ext_galaxy}
\end{figure*}

In the bottom panel of Figure~\ref{fig:multi_in_plane}, we continue to choose 
$(M_1, M_2)$ = (10, 10) $M_\odot$ and $\theta_3$ = 0.009 arc-sec. 
In addition, we fix $(\theta_1, \theta_2)$ = $(\frac{-2 \theta_3}{3}, \frac{2 \theta_3}{3})$. 
For reference, in red, we show the case with ($r_1, r_2$) = (1.0, 2.0) kpc. 
With the green and blue (black and yellow) curves, we vary $r_1$ ($r_2$) with 
$r_2$ ($r_1$) fixed at 2.0 (1.0) kpc. 
With the first two set of curves, we see that the rate of oscillation of 
the amplification factor increases (reduces) as $M_1$ moves away (closer) 
to the source. 
However, with $M_2$, we note the opposite: oscillations are more (less) rapid 
when $r_2$ is smaller (larger). 
Unlike in Figure~\ref{fig:multi_in_plane}, we see that the (radial) position of 
both the lenses can have a significant impact on the rate of oscillation. 
However, only the (radial) position of $M_1$ greatly effects the maximum value 
of amplification. 
This is as opposed to the top panel of Figure~\ref{fig:multi_in_plane}, where 
the (angular) position of both the lenses do have an effect on the
maximum value of amplification. 
We thus already begin to see the complex dependence of the amplification 
factor on the various parameters, wherein trends are not easy to extrapolate.

Finally, in Figure~\ref{fig:multi_off_plane}, we show the Off-Plane case. 
As mentioned above, this introduces three additional parameters: $\phi_1, 
\phi_2, \phi_3$, along with the In-Plane parameters, $\theta_1, \theta_2$. 
Although there are numerous ways in which these parameters can be varied, 
we choose the following scheme: in each panel, we consider the primary axis 
of the plane to be the line joining the source with a given object
(lens/observer).  
The second axis is considered to be the perpendicular dropped from one of 
the other two objects onto the primary axis. 
These two axes form the plane, and the third (remaining) object is displaced
out of this plane. 
In all plots, we fix $(r_1, r_2, r_3)$ = (1.0, 2.0, 3.0) kpc, $(M_1, M_2)$ = 
(10, 10) $M_\odot$. 
For simplicity, we define $\theta_a =$ 0.009 arc-sec and $\theta_b =
\frac{2 \theta_a}{3}$.  
Whenever not varied, we fix $(\theta_1, \phi_1)$ = $(\theta_2,
\phi_2)$ = (0, 0), and $(\theta_3, \phi_3)$ = $(\theta_a, 0)$ rad.

In the top panel, we consider the line joining the source with $M_1$ as the 
primary axis. 
For reference, in red, we show the case for which the two lenses are co-linear 
with the source. The green and blue curves consider the displacement of 
the observer from the plane. 
Both these curves oscillate faster than the reference, with the curve in 
green oscillating faster than the one in blue. 
However, the difference in rates of oscillation is not as large as one would 
expect from a comparison with Figure~\ref{fig:multi_in_plane}. 
The remaining three curves consider the displacement of $M_2$ out of the 
selected plane. 
With respect to the reference, the curve in black considers a displacement 
of $M_2$ away from the observer. 
As expected from previous Figures, oscillations grow more rapid, and the 
modulus curve oscillates to comparatively smaller values. 
The curve in yellow considers a displacement towards $M_2$ (albeit an 
`in-plane' displacement). 
In this case, we notice that oscillations grow slower and the modulus values 
oscillate to larger values, both of which are in agreement with
earlier observations. 
In the curve in pink, we reduce the angle of the observer, which essentially 
brings the observer closer to $M_2$. 
The observed trend is what one would expect from
Figure~\ref{fig:multi_in_plane}: oscillations begin to slow down, and
modulus values begin to increase. 

In the centre panel, we consider the line joining the source with $M_2$ as 
the primary axis. 
Akin to the previous case, the green and blue curves consider the displacement 
of the observer from the plane. 
Observations are qualitatively similar to the previous case: the green and blue 
curves oscillate faster than the reference, with the green curve oscillating 
slightly faster. 
Once again, the difference in rates of oscillation is smaller than expected. 
In the remaining three curves, $M_1$ is displaced off-plane. 
The black curve considers displacement of $M_1$ away from the observer. 
As opposed to previous observations, absolute values are similar to
the reference,  and the rates of oscillation are also comparable. 
However, if we consider an additional in-plane displacement (yellow curve), 
the modulus values begin to rise, but not as large as one would expect from 
Figure~\ref{fig:multi_in_plane}. 
Reducing $\theta_3$ (pink curve) seems to increase the modulus values, which 
is as expected, but the rate of oscillation drops significantly, which
contradicts Figure~\ref{fig:multi_in_plane}. 
We thus see that a combination of In-Plane and Off-Plane displacements lead to 
trends that are not always in agreement with In-Plane displacements alone, and 
the difference is larger for $M_1$.

In the bottom panel, we consider the line joining the source with the observer 
as the primary axis. 
Curves two - four (five - seven) consider off-plane displacements of
$M_1$ ($M_2$).   
The curve in green considers an off-plane displacement of $M_1$ away
from the observer.  
From Figure~\ref{fig:multi_in_plane}, one would expect the curve in green to 
oscillate faster than the one in red, but such a trend is not seen. 
However, we note that a slight reduction in the maximum value of
amplification, as expected. 
The curve in blue considers a displacement of $M_1$ towards the
observer, and as expected, this is accompanied by an increase in the
maximum value of amplification.  
The curve in black considers an in-plane displacement of $M_2$ towards
the source, and as expected from previous plots, the rate of
oscillation drops.  
However, we also note a very sharp increase in the maximum value of
amplification, which hasn't always been observed in earlier plots. 
The curves in yellow and pink consider displacements of $M_2$ away and
towards the observer, respectively. 
As expected, we note an increase and decrease in the rate of
oscillation, respectively.  
With the curve in cyan, we consider a displacement of $M_1$ towards
the observer, and in agreement with previous plots, we note an
increase in the maximum value of amplification.

\subsection{Effect of External Perturbations}
\label{ssec:2plane_galaxy}

So far, we have explored cases where both point mass lenses have been treated 
as isolated objects in their respective lens planes. 
While this is a starting point for lens systems within our own galaxy, a further 
generalization is required when lenses are embedded in host galaxies. 
In this subsection, we consider an extension of the special case discussed in
\S\ref{ssec:2plane_special}: scenarios wherein the two micro-lenses are collinear with 
the source, and the two micro-lenses are embedded in two distinct host galaxies. 
We assume that the effect of the host galaxy is constant across the Einstein radius
of the microlens. 
Under this framework, the following terms are introduced into the lensing potential 
(an extension of the single-plane case discussed in \citet{1992grle.book.....S}):

\begin{equation}
\begin{split}
    \psi_{\rm{ext}}\left(x_1,x_2,x_3,x_4\right) = \\ 
    \left(\frac{r_1 r_3}{2(r_3 - r_1)}\right)\left[\kappa_1 (x_1^2 + x_2^2) + \gamma_{11} (x_1^2 - x_2^2) + 2 \gamma_{12} x_1 x_2  \right] \\
    + \left(\frac{r_2 r_3}{2(r_3 - r_2)}\right)\left[\kappa_2 (x_3^2 + x_4^2) + \gamma_{21} (x_3^2 - x_4^2) + 2 \gamma_{22} x_3 x_4  \right],
\end{split}
\label{eq:ext_galaxy_lens_potl}
\end{equation}

where $\kappa_1$ ($\kappa_2$) and $(\gamma_{11}(\gamma_{21}), \gamma_{12}(\gamma_{22}))$  
correspond to the external convergence and shear parameters of the galaxy in which 
$M_1$ ($M_2$) is embedded. 
For simplicity, we choose $\gamma_{12}$ and $\gamma_{22}$ to be zero, which correspond to 
specific orientations of the two galaxies. Note that the various $x_i$'s in 
Equation~\ref{eq:ext_galaxy_lens_potl} correspond to angular co-ordinates.

We show the dependence of the amplification factor on the various external convergence/shear 
parameters in Figure~\ref{fig:multi_ext_galaxy}. 
For reference, in each of the panels, we show the isolated microlenses case in black. 
To ensure that numbers are meaningful with respect to galaxy lenses, we choose 
$(r_1, r_2, r_3)$ = $(100.0, 200.0, 300.0)$ Mpc. 
We fix $\theta_3$ = 0.00003 arc-sec and microlens masses are set to 10 $M_\odot$ each. 
For this choice of $\theta_3$, the curve in black is identical to the red curves from the 
previous figures.  

In the top panel, we consider cases with only external shear (i.e. $\kappa$'s are set to zero). 
When one/both of the two shear parameters is/are non-zero, we note the following: 
absolute value and phase of the amplification factor oscillate at different rates in comparison to the reference curve in black. 
In the centre panel, we consider cases with only external convergence. 
Unlike the previous case with only external shear, we notice a pattern in this case: 
`adding' external convergence into the lens system increases the magnitude of absolute values of the amplification factor at low frequencies ($\sim $10Hz).
In the bottom panel, we examine cases where both external convergence and shear are present 
in the lens system. 
One can see that the resultant curves with both non-zero shear and convergence look
like combinations of the above panels: the non-zero convergence leads to an increase in the overall magnitude
of the oscillations, while the non-zero shear introduces a modification in rate of oscillation.

As mentioned earlier, these observations depict the complex dependence
of the amplification factor on the various parameters. 
Unlike the single lens plane case (especially for the isolated point
mass lens), it isn't easy to make predictions on the behaviour of the
amplification factor when the values of multiple parameters are varied
simultaneously.  
From our observations, we recommend that the best way to study wave
effects in double- (or multi-, in general) plane lensing is to study
various possible lens systems case by case, instead of trying to
extrapolate results from a set of test cases. 
Apart from that, we also find that the Monte Carlo integrator provides
satisfactory results for the case of N = 2.  
However, for integrals corresponding to higher number of lens planes,
one may need to explore alternate methods
\citep[e.g.,][]{2020arXiv201003089F}.

\section{Conclusions}
\label{sec:conclusions}

In this work, we have explored the effects of wave optics in double
plane lensing.  
We have also compared the wave-optical effects of a double plane lens
with a single lens plane consisting of one and two-point mass lenses. 
We find that only for some arrangements one can identify similar
behavior in single and double plane lensing.  
For example, in some cases,  varying the observer position with
respect to the source yields a qualitatively similar behavior in both
single and double planes. 
However, apart from this, it is not straightforward to make
generalized statements about the behavior of lens systems: the number
of parameters is just too large, even for the relatively simple case
of two lens planes.  

One would expect the complexity to further increase for a large number
of lens planes.  
However, the probability of observing lens system with more than two lens
planes is significantly lower.
The increase in complexity also holds for double plane lenses if we increase
the number of point mass lenses in one or both of the lens planes.
The presence of galaxy lenses in such systems would introduce an
additional shear and convergence, which is expected to further complicate the analysis. 
Keeping this in mind, we conclude that the best way to study wave effects in
multi-plane lensing is to tackle separate lens systems case by case instead of 
hoping to make predictions based on a few test cases.

\section*{Acknowledgements}

RR would like to thank the Department of Science and Technology,
Government of India for being awarded the INSPIRE scholarship. 
AKM would like to thank Council of Scientific and Industrial Research
(CSIR) India for financial support through research fellowship
No. 524007.  
Authors acknowledge the use of IISER Mohali HPC facility. Authors would like to thank Kandaswamy Subramanian for valuable comments on the manuscript. This research has made use of NASA's Astrophysics Data System
Bibliographic Services. 

\bibliography{references}
\bibliographystyle{apj}

\appendix
\section{Geometric Optics}
\label{sec:appendix_A}

For greater insights into Figure~\ref{fig:binary_lens_wave}, we here show 
the critical lines and caustics corresponding to the various cases discussed earlier. 
When the source happens to be outside the caustic, three images of the source are formed. 
However, when the source moves into the caustic, an additional two images are 
produced, thereby increasing the complexity of the amplification factor.

\begin{figure*}[h!]
    \centering
	\includegraphics[width=16.6cm, height=6cm]{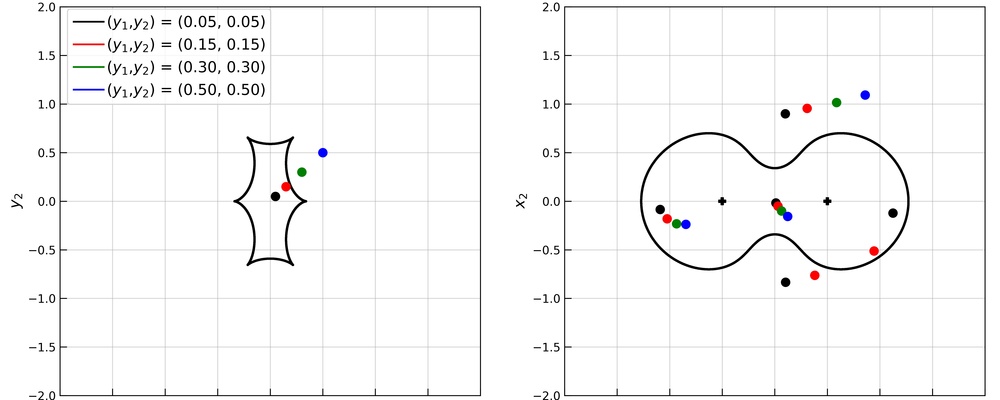}
	\includegraphics[width=16.6cm, height=6cm]{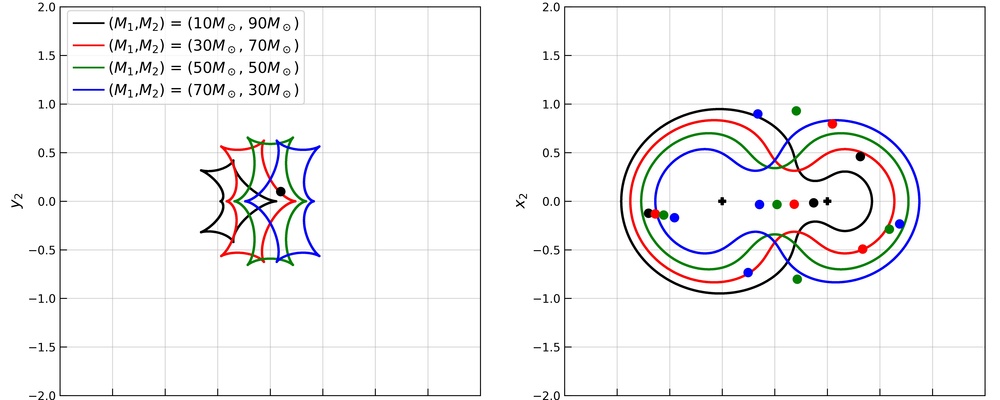}
	\includegraphics[width=16.6cm, height=6.5cm]{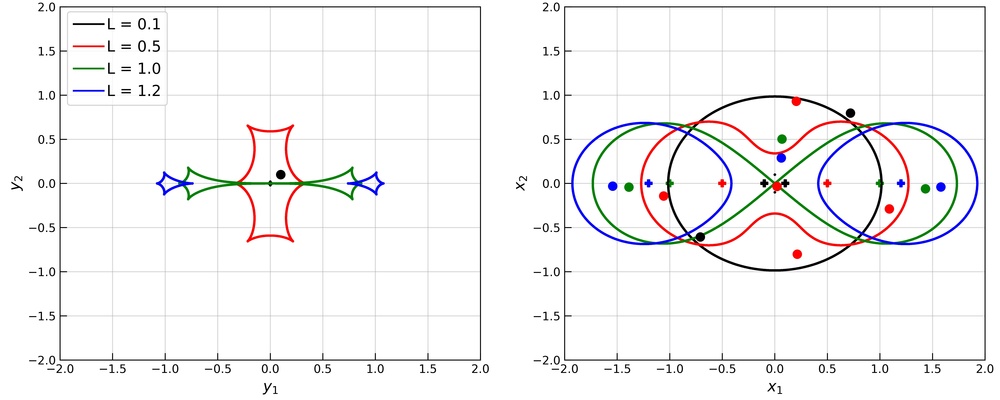}
	\caption{Caustics (left) and Critical lines (right) corresponding 
	to the cases discussed in Figure \ref{fig:binary_lens_wave}. The 
	various filled circles correspond to the position of the source 
	(images) in the left (right) panel. When the source happens to be 
	outside (inside) the caustic, three (five) images are formed.}
	\label{fig:binary_lens_geom}
\end{figure*}

\end{document}